\documentclass[letterpaper]{article} 
\usepackage[preprint]{aaai2027}  
\usepackage[hyphens]{url}  
\usepackage{graphicx} 
\usepackage{natbib}  
\usepackage{caption} 
\usepackage{algorithm}
\usepackage{algorithmic}
\usepackage{booktabs}
\usepackage{multirow}
\usepackage{amsmath,amssymb,amsfonts}
\usepackage{textcomp}
\usepackage{xcolor}
\usepackage{enumitem}
\title{An LLM-Assisted Multi-Objective AutoML Framework for Efficient Network Intrusion Detection}

\author{
    Ailin Barzegar\textsuperscript{\rm 1},
    Khalil El-Khatib\textsuperscript{\rm 1},
    Li Yang\textsuperscript{\rm 1}
}
\affiliations{
    \textsuperscript{\rm 1}Faculty of Business and Information Technology, Ontario Tech University\\
    Oshawa, ON, Canada\\
    ailin.barzegar@ontariotechu.ca; khalil.el-khatib@ontariotechu.ca; li.yang@ontariotechu.ca
}

\begin{document}

\maketitle

\begin{abstract}
Conventional Machine Learning (ML)-based Intrusion Detection 
Systems (IDSs) require manual design, optimize only detection 
performance, and neglect inference latency and model size, which 
are critical for resource-constrained Internet of Things (IoT) 
and Electric Vehicle Charging System (EVCS) deployments. 
Automated ML (AutoML) reduces manual effort but explores broad 
search spaces inefficiently. Meanwhile, Multi-Objective 
Optimization (MOO) methods lack guided search initialization, 
leading to long search times. This paper proposes a Large Language 
Model-Assisted Multi-Objective Automated ML (LLM-MOO-AutoML) 
framework for efficient intrusion detection. A local Llama 3.1 8B 
model analyzes a compact dataset profile and generates a bounded, 
validated policy for automated data balancing (AutoDP) and 
feature engineering (AutoFE). This policy narrows the search 
space for Non-dominated Sorting Genetic Algorithm III (NSGA-III), 
which jointly optimizes the LightGBM feature-selection threshold 
and key hyperparameters across three objectives: maximizing 
weighted F1-score, minimizing 99th-percentile (P99) inference 
latency ratio, and minimizing model size ratio. Experiments on 
CICIDS2017, IoTID20, and CICEVSE2024 demonstrate that the 
proposed framework achieves the highest weighted F1-scores on CICIDS2017 (99.788\%), IoTID20 (99.778\%), and CICEVSE2024 (99.989\%), with feature sets reduced to as few as 10 features, P99 latency as low as 0.00038~ms, and model size 
as small as 0.1073~MB. The LLM-guided configuration reduced NSGA-III search time relative to unconstrained multi-objective search on all datasets at the largest trial budget and in most comparisons at smaller budgets, indicating that dataset-aware search-space reduction supports efficient, deployment-aware intrusion detection.
\end{abstract}

\section{Introduction}

The rapid proliferation of Internet of Things (IoT) devices and Electric
Vehicle Charging Systems (EVCSs) has significantly expanded the attack surface
of modern networks. By 2026, the number of connected IoT devices is projected
to exceed 26~billion globally~\citep{sharma2026iot}, spanning smart home
sensors, industrial controllers, medical monitors, and vehicle telematics.
EVCSs represent a particularly critical subset: connected chargers communicate
with grid management systems, payment processors, and vehicle on-board units
over open network protocols, creating multiple entry points for adversaries.
A successful cyberattack on EVCS infrastructure can disrupt charging
availability, manipulate billing, overload grid substations, or compromise
driver safety~\citep{dehrouyeh2024tinyml}.
Connected chargers, smart grid interfaces, and IoT sensors enable richer
services but also expose critical infrastructure to Denial-of-Service (DoS),
scanning, Man-in-the-Middle (MITM), and other network
attacks~\citep{sharma2026iot,dehrouyeh2024tinyml}.
Intrusion Detection Systems (IDSs) are a primary defense: they monitor network
traffic continuously and flag malicious activity that bypasses preventive
controls such as firewalls and encryption.

Machine Learning (ML)-based IDSs have become the dominant paradigm because ML models learn
complex, nonlinear relationships between network-flow features and attack
classes~\citep{yang2024automl_zero_touch}. However, developing an effective
IDS remains labor-intensive.  Automated ML (AutoML) reduces this burden by automating
Automated Data Pre-processing (AutoDP), Automated Feature Engineering
(AutoFE), and Combined Algorithm Selection and Hyperparameter Optimization
(CASH)~\citep{yang2024autonomous_cyber}.

Despite these advances, two complementary limitations persist.
First, conventional AutoML methods optimize a single objective, typically
F1-score, without explicitly accounting for inference latency or model size.
For EVCS and IoT deployments where edge devices operate under strict memory
and real-time constraints, an accurate but slow or large model is not
deployable~\citep{yang2026moo_globecom}.
 Multi-Objective Optimization (MOO) frameworks such as Non-dominated Sorting Genetic Algorithm III (NSGA-III) address this by producing a Pareto front of
solutions that trade off detection performance, latency, and model compactness.
Second, MOO over unconstrained hyperparameter spaces can be computationally
expensive, spending many evaluations in poor regions.
Large Language Models (LLMs) have recently been shown to improve AutoML
efficiency by generating compact, dataset-aware policies that bound the search
space before optimization begins~\citep{yang2026llm_automl}.
This paper addresses both limitations simultaneously 
by using a validated LLM policy to constrain and 
initialize NSGA-III, empirically demonstrating that 
dataset-aware bounds improve both solution quality 
and shorten search time under a fixed trial budget.
This paper proposes LLM-MOO-AutoML, which combines 
two complementary components: (i) an LLM policy 
generation stage that analyzes a compact dataset 
profile and produces bounded, validated AutoDP and 
AutoFE configurations together with explicit 
hyperparameter ranges and feature-selection threshold 
bounds that seed the NSGA-III population; and (ii) an 
NSGA-III multi-objective search stage that uses those 
LLM-generated configurations to jointly optimize the 
LightGBM feature-selection threshold and hyperparameters 
under three deployment-aware objectives: weighted 
F1-score, P99 inference latency ratio, and model size 
ratio.
The proposed LLM-MOO-AutoML framework is designed
to produce IDS models suitable for deployment in
resource-constrained environments typical of IoT and
EVCS infrastructure, by jointly optimizing detection
accuracy, inference latency, and model size.
Concretely, the framework targets three deployment
contexts: (i) EVCS network monitoring, where low P99
inference latency and small model size are critical
for real-time detection of DoS and reconnaissance
attacks; (ii) IoT network security, where compact
models with few selected features reduce memory
footprint and inference cost on resource-limited
devices; and (iii) general enterprise network
security, where the AutoML pipeline can be retrained
periodically on new traffic captures without manual
feature engineering. In all three cases, the AutoML
pipeline runs offline and only the final LightGBM
model is used for inference, keeping deployment
lightweight. Validation is performed on standard
benchmark datasets under controlled experimental
conditions.
The main contributions are:
\begin{enumerate}
  \item We propose LLM-MOO-AutoML, a novel framework that 
integrates LLM-generated, dataset-aware AutoDP and AutoFE 
policies with NSGA-III multi-objective optimization for 
deployment-aware IDS. The LLM policy is validated and 
converted into bounded NSGA-III search spaces, preventing 
invalid configurations and focusing the search on 
dataset-relevant regions before any candidate is evaluated.
 \item We design the first integration of LLM-guided AutoDP and 
AutoFE with a three-objective NSGA-III search jointly optimizing 
weighted F1-score, P99 inference latency ratio, and model size 
ratio. Under a fixed trial budget, the LLM-constrained search 
consistently finds higher-quality Pareto solutions than 
unconstrained search, achieving P99 latency as low as 
0.00038~ms, model size as small as 0.1073~MB, and feature sets 
as compact as 10 features across three benchmark datasets.
\item We empirically demonstrate that LLM-guided search-space bounds reduce NSGA-III optimization search time by 34.0-70.5\% across CICIDS2017, IoTID20, and CICEVSE2024 at the 50-trial budget, and in most comparisons at smaller budgets, while maintaining equivalent or superior solution quality compared with unconstrained multi-objective search. This confirms that dataset-aware bounds prevent wasted evaluations in unproductive hyperparameter regions, improving both search time and deployment efficiency.
\end{enumerate}
The remainder of this paper is organized as follows. 
Section~2 reviews related work on ML-based, AutoML-based, 
MOO-based, and LLM-assisted IDS methods. Section~3 
describes the proposed LLM-MOO-AutoML framework. 
Section~4 presents experimental results, search 
efficiency analysis, and discussion. Section~5 concludes 
the paper.

\section{Related Work}
\label{sec:related}

We review four interconnected lines of research that motivate the
proposed framework: (A)~ML-based and optimized IDSs,
(B)~AutoML-based IDSs, (C)~MOO-based IDSs, and
(D)~LLM-assisted ML and AutoML.

\subsection{ML-based and Optimized IDSs}

ML-based IDSs have become the dominant approach for network intrusion
detection because ML models learn complex, nonlinear relationships
between network-flow features and attack
classes~\citep{yang2024automl_zero_touch,fatani2021iot}.
Among model families, tree-based ensembles are consistently competitive
with or superior to deep learning on tabular network-traffic
data~\citep{grinsztajn2022tree}, making them a natural backbone for
efficient IDS development.

\citet{jin2020lightgbm} proposed SwiftIDS, a real-time IDS
that exploits LightGBM's histogram-based learning and a parallel
detection mechanism to achieve low per-sample inference cost on
large-scale traffic data.
\citet{gutierrez2023iotids} proposed an
IDS based on Random Forest and Extra Trees evaluated on a new IoT
communication dataset, demonstrating strong generalisation across
heterogeneous device types.
\citet{yang2022mth} proposed a multi-tiered hybrid
IDS for Internet of Vehicles that stacks tree-based learners with a
meta-classifier, achieving high detection rates across diverse attack
categories with low retraining overhead.
\citet{ferrag2020deep} conducted a comprehensive benchmarking
study of deep learning architectures for network IDS across multiple
public datasets, finding that simpler ensemble methods often match deep
models while requiring less tuning and computational cost.

In terms of optimized and metaheuristic IDSs,
\citet{elmasry2020pso} used double Particle Swarm Optimization (PSO) to evolve the
depth, units, and dropout of Long Short-Term Memory (LSTM) networks for network IDS, reducing
the manual effort of deep architecture design.
\citet{naeem2025gacnn} combined Genetic Algorithm (GA) search with a Convolutional Neural Network (CNN)
transfer-learning backbone for multiclass vehicle-network security
classification, showing that metaheuristic wrappers can substantially
improve deep-learning IDS performance beyond manual design.
\citet{khraisat2019survey} surveyed hybrid IDS designs
combining signature-based rules with ML anomaly detectors, identifying
that hybrid approaches reduce false-positive rates compared with
pure ML methods at the cost of higher maintenance overhead.
\citet{mishra2020deep} proposed a deep learning IDS for
fog-based IoT environments that adapts to changing traffic distributions
using incremental retraining, addressing the concept-drift limitation
of static classifiers.

For IoT and EVCS-specific IDSs,
\citet{diro2018distributed} proposed a distributed
deep-learning IDS for fog-computing IoT environments that uses
parameter sharing to reduce per-device training cost.
\citet{ullah2020iotid20} introduced the IoTID20 dataset
and an ML IDS evaluation framework, enabling standardised comparison
of IoT intrusion detection methods across Mirai, DoS, scanning, and
MITM attack categories.

Despite strong detection performance, these methods rely on manual
feature selection, model design, or objective definition.
They do not automate the full ML pipeline and rarely consider inference
latency or model size as explicit optimization objectives.

\subsection{AutoML-based IDSs}

AutoML frameworks automate key ML pipeline stages, data preprocessing,
feature engineering, model selection, and hyperparameter tuning,
reducing manual effort and design bias~\citep{yang2022iot_analytics}.

\citet{yang2024autonomous_cyber} proposed an autonomous
AutoML IDS that jointly performs AutoDP, AutoFE, and CASH via Bayesian
optimization with Tree-structured Parzen Estimator (TPE), achieving competitive F1-scores on CICIDS2017
and IoTID20.
\citet{singh2022automlid} introduced AutoML-ID, which
applies automated model selection and Hyperparameter Optimization to wireless sensor network
IDS, reducing development time while maintaining competitive accuracy.
\citet{khan2023oeids} proposed Optimized Ensemble (OE)-IDS, an AutoML soft-voting
ensemble aggregating LightGBM, XGBoost, and Random Forest, showing
that ensemble diversity found through automated search improves
detection precision.
\citet{ferrag2022edge} introduced Edge-IIoTset, a
comprehensive IoT and IIoT cybersecurity dataset paired with an
AutoML IDS evaluation, showing that automated pipelines can be adapted
to edge-constrained environments.

In terms of automated feature engineering for IDS,
\citet{yang2024automl_zero_touch} provided a systematic
analysis of AutoML components for zero-touch network security,
identifying automated feature selection as the highest-value step for
tabular traffic data.
\citet{kasongo2020performance} evaluated wrapper-based
feature selection methods for LightGBM on UNSW-NB15 dataset, showing that
removing low-importance features reduces inference time by up to 30\%
with negligible F1 degradation.

\subsection{Multi-Objective Optimization-based IDSs}

MOO methods produce Pareto-optimal solution sets that simultaneously
balance competing objectives, making them well-suited for
deployment-aware IDS development where detection accuracy, inference
speed, and memory footprint must all be controlled~\citep{deb2014nsga3}.

Regarding MOO for feature selection in IDS,
\citet{subramani2023mopso} applied Multi-Objective Particle Swarm Optimization (MOPSO) to
feature selection for IoT wireless sensor network IDS, simultaneously
minimising feature count and maximising detection accuracy.
\citet{moustafa2021federated} integrated MOO into a
federated learning IDS for IoT, optimising detection rate and
communication overhead simultaneously.
\citet{zhou2021multi} proposed a multi-objective feature
selection method for network IDS using a decomposition-based
evolutionary algorithm, demonstrating Pareto-optimal trade-offs
between feature dimensionality and classification accuracy on several
benchmark datasets.

To have MOO for joint model and feature optimization,
\citet{yang2025moo_journal} proposed a MOO-AutoML IDS
using NSGA-III to jointly optimise LightGBM hyperparameters and
feature-selection thresholds under three objectives: weighted F1-score,
P99 inference latency ratio, and model size ratio.
This work established the three-objective formulation and baseline
results adopted in the proposed framework.
\citet{yang2026moo_globecom} extended this to EVCS cybersecurity
on CICEVSE2024, achieving near state-of-the-art F1-scores while
reducing P99 latency by up to 49\% and model size by up to 70\%
compared with single-objective baselines.

\subsection{LLM-assisted ML and AutoML}

LLMs have demonstrated the ability to encode broad domain knowledge
about ML pipelines, data characteristics, and algorithm
behavior~\citep{tornede2024automl_llm,gu2025llm_survey}.
This knowledge can warm-start, guide, or bound automated search
procedures, reducing the number of expensive model evaluations
required. \citet{hollmann2023caafe} proposed CAAFE, which
iteratively generates context-aware feature transformations for
tabular data using an LLM prompted with column names and dataset
descriptions, producing semantically meaningful features that improve
downstream accuracy without exhaustive search.

In terms of LLM-assisted hyperparameter optimization,
\citet{liu2024llambo} proposed Large Language Model-based Bayesian Optimization (LLAMBO), which integrates LLM
guidance into Bayesian optimization for HPO.
\citet{luo2024llm_automl} evaluated LLMs as AutoML
assistants across model selection, feature engineering, and HPO,
finding that structured prompts with dataset statistics consistently
outperform unstructured queries on a broad range of tasks.

For LLM-assisted AutoML for IDS,
\citet{yang2026llm_automl} proposed the first LLM-assisted AutoML
framework for IoT IDS.
A local Llama~3.1~8B model receives a compact dataset profile and
generates a bounded, validated policy for AutoDP, AutoFE, and CASH
(BO-TPE).
\citet{li2024idsagent} proposed IDS-Agent, an LLM-driven
agent with specialised IDS tools, memory, and external knowledge
that reasons over network traffic and produces explainable intrusion
decisions.
\citet{ali2026huntgpt} introduced HuntGPT, which combines
a Random Forest anomaly detector with explainable AI components and an
LLM-based conversational interface, enabling analysts to query
detected threats in natural language.

Current LLM-assisted AutoML methods for IDS optimise a single objective
and do not combine LLM guidance with MOO-based deployment-aware search.
The proposed LLM-MOO-AutoML framework is the first to address this gap
by using LLM-generated AutoDP and AutoFE policies to constrain and
initialize an NSGA-III search that jointly optimises detection
performance, P99 inference latency, and model compactness across
general networks, IoT, and EVCS environments.
In summary, the reviewed literature reveals four interconnected gaps that
collectively motivate the proposed framework. Traditional ML-based IDSs,
including tree-ensemble methods such as SwiftIDS~\citep{jin2020lightgbm}
and MTH-IDS~\citep{yang2022mth}, and deep learning benchmarks~\citep{ferrag2020deep},
deliver strong detection accuracy but require manual feature engineering,
model selection, and tuning, with no consideration of inference latency or
model size. AutoML-based IDSs~\citep{yang2024autonomous_cyber,khan2023oeids,
singh2022automlid,ferrag2022edge} reduce this manual burden through automated
CASH and feature selection pipelines, yet they optimize a single objective
over broad, dataset-agnostic search spaces that do not account for the
memory and latency constraints of IoT and EVCS edge deployments.
MOO-based IDSs~\citep{yang2025moo_journal,yang2026moo_globecom,
subramani2023mopso} address deployment efficiency by producing
Pareto-optimal solutions that trade off detection performance, inference
latency, and model compactness, but rely on generic population initialisation
over unconstrained search spaces, leading to long search times and costly
evolutionary evaluations. LLM-assisted AutoML
methods~\citep{yang2026llm_automl,hollmann2023caafe,liu2024llambo} improve
search efficiency by encoding dataset-aware priors into bounded pipeline
configurations, but remain limited to single-objective optimization and have
not been combined with MOO-based deployment-aware search.
The proposed LLM-MOO-AutoML framework is the first to bridge all four gaps
simultaneously: a local Llama~3.1~8B model generates validated AutoDP and
AutoFE policies from a compact dataset profile, narrowing the NSGA-III search
space to dataset-aware, deployment-relevant regions before any candidate is
evaluated, while NSGA-III jointly optimises weighted F1-score, P99 inference
latency ratio, and model size ratio across general network, IoT, and EVCS
environments. This combination of LLM-guided search-space reduction and
multi-objective deployment-aware optimization is novel and distinguishes the
proposed framework from all prior work reviewed above.
\section{Proposed LLM-MOO-AutoML Framework}
\label{sec:framework}

\subsection{System Overview}
\begin{figure*}[!t]
  \centering
  \includegraphics[width=2\columnwidth]{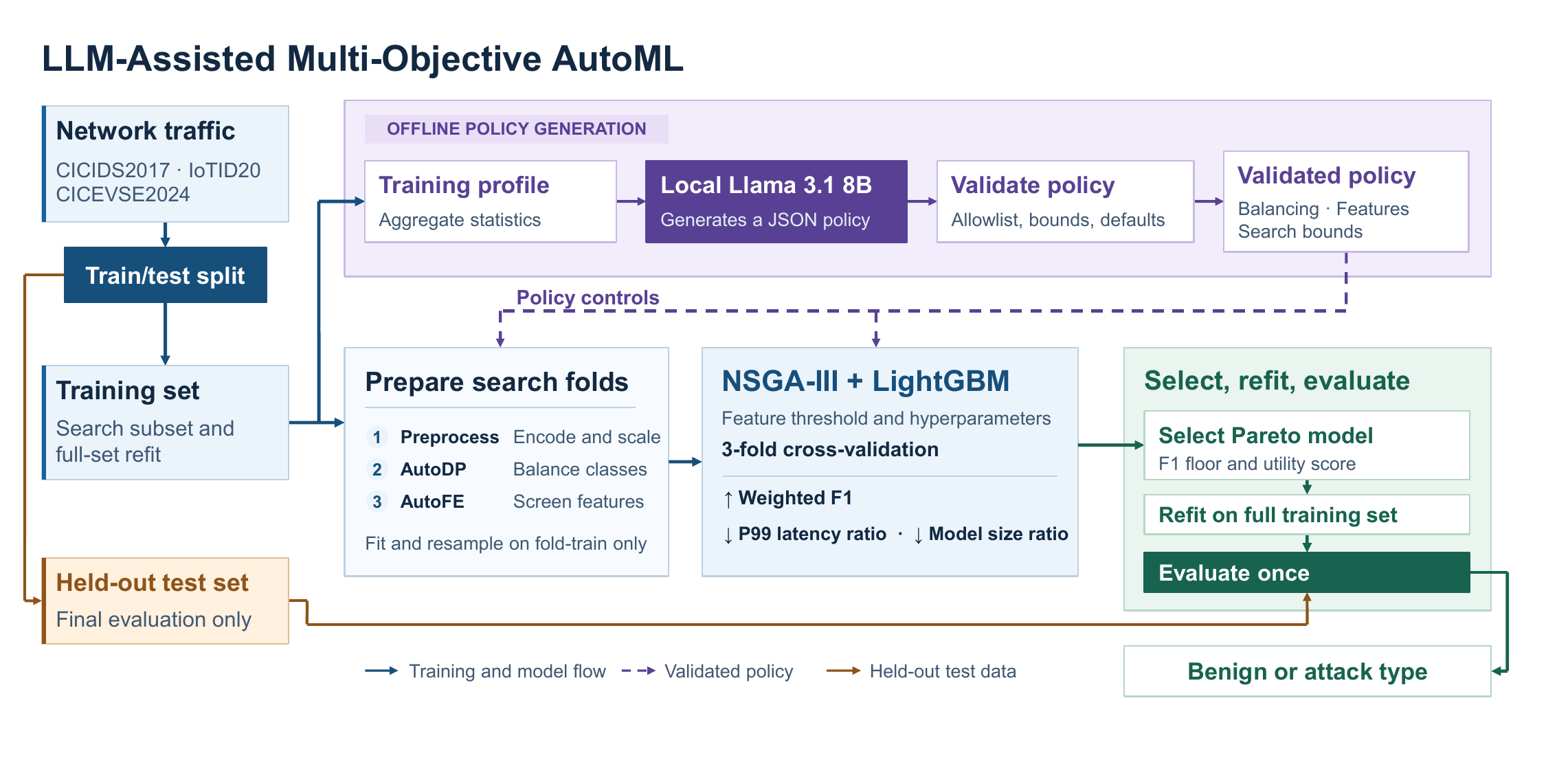}
  \caption{Overview of the proposed LLM-MOO-AutoML IDS framework.}
  \label{fig:framework}
\end{figure*}

The proposed framework operates in three coordinated stages illustrated in Figure~\ref{fig:framework}.
\textbf{Stage~1:} A compact dataset profile $\mathcal{P}$ is constructed from the outer training partition and provided to a local Llama~3.1~8B model ~\citep{llama3}, which returns a validated AutoDP and AutoFE policy together with LightGBM hyperparameter bounds for the MOO search.
\textbf{Stage~2:} The validated policy controls data preparation on each cross-validation fold: label encoding, $z$-score scaling, AutoDP class balancing, and AutoFE feature screening.
\textbf{Stage~3:} NSGA-III samples the feature-selection threshold $\tau$ within the LLM-specified bounds and searches the hyperparameter space $\Theta$ within the LLM-guided ranges; the deployment-oriented Pareto solution is then selected, refitted on the full training set, and evaluated once on the held-out test set.
Formally, let $\mathcal{D}=\{(\mathbf{x}_i,y_i)\}_{i=1}^n$ be a labeled
traffic dataset, where $n$ is the number of training samples,
$\mathbf{x}_i$ is a $d$-dimensional feature vector of
network-flow statistics, and $y_i$ is the traffic class
label.
The LLM-guided policy is:
\begin{equation}
  \pi = \{\beta,\,\phi,\,\mathcal{T},\,\Theta\}\in\Pi,
  \label{eq:policy}
\end{equation}
where $\beta=\{b,q,r_{\max}\}$ is the set of AutoDP data-balancing
controls (oversampling method, target quantile, and growth cap),
$\phi=\{\tau_{\mathrm{corr}}, \gamma_{\mathrm{imp}}, m_{\mathrm{imp}},
k_{\mathrm{MI}}, m_{\min}\}$ is the set of AutoFE feature-selection
controls (correlation threshold, importance threshold, feature cap,
Mutual Information (MI) prefilter width, and minimum feature count),
$\mathcal{T}=[\tau_{\mathrm{lo}},\tau_{\mathrm{hi}}]$ is the
LLM-bounded sampling range for the accumulated importance threshold
$\tau$, $\Theta$ is the LLM-bounded LightGBM hyperparameter search
space, and $\Pi$ is the admissible policy space enforced by the
validation layer.
NSGA-III solves:
\begin{equation}
  \min_{\tau\in\mathcal{T},\,\theta\in\Theta}
  \mathbf{g}(\tau,\theta)=
  \bigl[1-F_1^w,\;R_{P99},\;R_{\mathrm{size}}\bigr],
  \label{eq:moo}
\end{equation}
where $F_1^w$ is the mean weighted F1-score under stratified $K{=}3$
cross-validation folds (measuring detection performance across all
attack classes weighted by class support), $R_{P99}$ is the ratio of
the candidate model's 99th-percentile per-sample inference latency to
that of a full-feature LightGBM baseline (measuring tail inference
speed), and $R_{\mathrm{size}}$ is the ratio of the candidate model's
serialised byte size to that of the same baseline (measuring memory
footprint for edge deployment).
Minimising $1-F_1^w$ maximises detection performance.
The three objectives conflict: more features and deeper trees improve
$F_1^w$ but increase $R_{P99}$ and $R_{\mathrm{size}}$; the LLM
bounds narrow the search to regions where a favorable trade-off is
more likely.

\subsection{Stage 1- LLM Policy Generation and Validation}

The dataset profile
$\mathcal{P}=\{n,\,d,\,d_{\mathrm{num}},\,d_{\mathrm{cat}},\,
\mathbf{c}_y\}$
is computed from the outer training partition only, where $d$ is the
total feature count, $d_{\mathrm{num}}$ and $d_{\mathrm{cat}}$ are the
counts of numerical and categorical features, and
$\mathbf{c}_y$ is the class-count
vector (the number of samples per class, which characterises imbalance
severity).
Only aggregate statistics are included; raw traffic records are never
sent to the LLM.

A local Llama~3.1~8B model receives $\mathcal{P}$ with a fixed JSON
schema at temperature~0 and returns policy $G$ with three groups of
parameters:

\begin{itemize}
  \item \textbf{AutoDP controls} $\beta=\{b,q,r_{\max}\}$:
        $b \in \{\text{SMOTE, ADASYN, random, none}\}$ is the
        oversampling algorithm applied to minority classes;
        $q$ is the quantile of the class-count
        distribution used as the shared resampling target, where a
        lower $q$ produces lighter balancing;
        $r_{\max} $ is the per-class growth cap that
        limits how many times a minority class may grow relative to
        its original size, preventing excessive synthetic sample
        generation for very rare attack types.

  \item \textbf{AutoFE controls} $\phi$:
        $k_{\mathrm{MI}} \in [10,\,d]$ is the number of top features
        retained by the mutual-information pre-screen before the more
        expensive LightGBM ranking;
        $\gamma_{\mathrm{imp}}$ is the minimum
        fraction of total LightGBM gain importance that the selected
        feature subset must accumulate (a higher value retains more
        features);
        $m_{\mathrm{imp}} \in [10,\,d]$ is the hard upper bound on
        the number of features kept after importance selection;
        $\tau_{\mathrm{corr}}$ is the maximum
        absolute Pearson correlation allowed between any two retained
        features, above which the less important one is removed;
        $m_{\min}$ is the minimum number of features
        that must survive all AutoFE steps, preventing the feature
        set from collapsing to an unusable size.

  \item \textbf{MOO search bounds} $\{\mathcal{T},\Theta\}$:
        $[\tau_{\mathrm{lo}},\tau_{\mathrm{hi}}]$
        is the interval within which NSGA-III samples the accumulated
        importance threshold $\tau$; a narrower interval focuses the
        search on a promising feature-count region.

\end{itemize}

The generated policy $G$ is never executed directly.
A validation layer converts it to executable $\pi$ by:
(i)~checking $b$ against the allowlist and replacing invalid values;
(ii)~clipping $q$ to $[0.50,0.75]$ and $r_{\max}$ to $[2.0,10.0]$;
(iii)~clipping all AutoFE controls to their safe ranges and
lower-bounding $m_{\min}$ at 5;
(iv)~intersecting $[\tau_{\mathrm{lo}},\tau_{\mathrm{hi}}]$ with
$[0.50,1.00]$, defaulting to $[0.75,1.00]$ on failure;
(v)~intersecting each LightGBM range with its safe envelope, defaulting
to the full envelope on failure.
These rules ensure $\pi\in\Pi$ regardless of LLM output quality.

 For CICIDS2017 ($n$=48,569, $d$=77, 
15 classes, $\rho$=295), Llama~3.1~8B returned: 
balancer=SMOTE, $q$=0.60, $r_{\max}$=6.0 (AutoDP); 
$k_{\mathrm{MI}}$=70, $m_{\mathrm{imp}}$=60, 
$\gamma_{\mathrm{imp}}$=0.985 (AutoFE); 
$\tau \in [0.80, 0.97]$, $n\_\text{estimators} \in [50, 180]$ 
(MOO bounds). Compared with the unconstrained default 
($\tau \in [0.50, 1.00]$, all 77 features eligible), the LLM policy caps the retained feature set at 60 features, sets lighter minority-class balancing appropriate for the 80.2\% BENIGN majority, and narrows the NSGA-III search space, yielding a 24-feature final model after importance selection and Pearson pruning (vs. 33 for MOO-AutoML (Yang 2026b)) and reducing optimization time by 37.6\% (3535.0 s to 2206.4 s).

\subsection{Stage 2- LLM-Guided AutoDP and AutoFE}

Before balancing and feature selection, categorical traffic features are
label-encoded (each category mapped to an integer) and all numerical
features are z-score standardized:
\begin{equation}
  x' = \frac{x - \hat{\mu}_{\mathrm{tr}}}{\hat{\sigma}_{\mathrm{tr}}},
  \label{eq:zscore}
\end{equation}
where $\hat{\mu}_{\mathrm{tr}}$ and $\hat{\sigma}_{\mathrm{tr}}$ are the
mean and standard deviation of the feature estimated on the fold-training
partition, and $x'$ is the standardised value applied identically to the
paired validation or held-out partition using the same statistics.
Standardisation prevents features with large numerical ranges from
dominating distance-based oversampling.

\textbf{AutoDP.}
The imbalance ratio
\begin{equation}
  \rho = \frac{\max(\mathbf{c})}{\min(\mathbf{c})}
  \label{eq:imbalance}
\end{equation}
measures how many times more samples the largest class has than the
smallest; $\rho > 1.5$ triggers partial resampling.
The common target count for minority classes is
\begin{equation}
  t_q = \min\!\bigl(\max(\mathbf{c}),\;
        \max\!\bigl(Q_q(\mathbf{c}),\;\min(\mathbf{c})+1\bigr)\bigr),
  \label{eq:tq}
\end{equation}
where $Q_q(\mathbf{c})$ is the $q$-quantile of the class-count
distribution, i.e.\ the count value below which $q$ fraction of all
class sizes fall.
The per-class oversampling target applies the growth cap:
\begin{equation}
  t_j = \min\!\bigl(t_q,\;\lfloor r_{\max} c_j \rfloor\bigr),
  \label{eq:autodp}
\end{equation}
where $c_j$ is the current sample count of class $j$; class $j$ is
oversampled only when $c_j < t_j$.
SMOTE~\citep{chawla2002smote} generates synthetic samples by
interpolating between $k$-nearest minority neighbours;
ADASYN~\citep{he2008adasyn} concentrates generation near
decision-boundary regions; random oversampling duplicates existing
samples and serves as the fallback when synthetic sampling fails.
All resampling is applied to fold-training data only.

\textbf{AutoFE.}
Four sequential steps reduce the feature space, all fitted on
fold-training data:
(1)~\emph{Near-constant removal}: features with variance below
$\epsilon_{\mathrm{var}}=10^{-8}$ are discarded as they carry no
discriminative signal;
(2)~\emph{MI prefiltering}: the top-$k_{\mathrm{MI}}$ features ranked
by mutual information with the class label are retained, where MI
measures how much a feature reduces uncertainty about the class;
(3)~\emph{Gain-importance selection}: a reference LightGBM model (100
estimators) ranks features by gain importance $I_j$, the total
reduction in training loss contributed by feature $j$ across all tree
splits, normalized so $\sum_j I_j=1$, and the smallest feature subset
$S_\tau$ satisfying
\begin{equation}
  S_\tau=\{f_1,\ldots,f_m\};\quad
  \frac{\sum_{i=1}^{m}I_i}{\sum_{j=1}^{d}I_j}\geq\tau,
  \label{eq:autofs}
\end{equation}
is retained, bounded by $[m_{\min},\,m_{\mathrm{imp}}]$;
(4)~\emph{Pearson pruning}: for any two retained features with absolute
correlation
\begin{equation}
  R_{pq}=|\mathrm{corr}(x_p,x_q)|>\tau_{\mathrm{corr}},
  \label{eq:corr}
\end{equation}
the feature with the lower $I_j$ is removed; $m_{\min}$ features are
restored if pruning reduces the set below the floor.

\subsection{Stage 3- NSGA-III Multi-Objective Search}
\label{subsec:stage3}

Each NSGA-III candidate $(\tau,\theta)$ uses $S_\tau$
(Eq.~(\ref{eq:autofs})) as its feature subset and trains a LightGBM
model with hyperparameter vector $\theta$.

The three objectives evaluated per candidate are:
\begin{equation}
  R_{P99}=\frac{L^{\mathrm{cand}}_{P99}}{L^{\mathrm{base}}_{P99}},
  \qquad
  R_{\mathrm{size}}=\frac{S^{\mathrm{cand}}}{S^{\mathrm{base}}},
  \label{eq:ratios}
\end{equation}
where $L^{\mathrm{cand}}_{P99}$ is the candidate's 99th-percentile
per-sample prediction time (capturing worst-case tail latency rather
than the mean), $L^{\mathrm{base}}_{P99}$ is the same metric for the
full-feature baseline, $S^{\mathrm{cand}}$ is the candidate model's
serialized byte size, and $S^{\mathrm{base}}$ is the baseline's serialised
size; a ratio below~1 means the candidate is more efficient than the
baseline.
NSGA-III~\citep{deb2014nsga3} ranks all $T{=}50$ candidate solutions
into successive non-dominated fronts and uses reference-point-based
selection to preserve diverse trade-offs across the three objectives.
The Pareto front quality is assessed by hypervolume (HV), the volume
of objective space dominated by the front relative to a reference point.

After $T$ trials, a deployment-oriented solution is selected from the
Pareto front using the utility score:
\begin{equation}
  u(\tau,\theta)=w_{F_1}(1-g_1)-w_{P99}\,g_2-w_{\mathrm{size}}\,g_3,
  \label{eq:utility}
\end{equation}
where $(g_1,g_2,g_3)=\mathbf{g}(\tau,\theta)$ are the three objective
values; $w_{F_1}{=}0.95$ reflects the primary importance of detection
accuracy for IDS, while $w_{P99}{=}0.03$ and
$w_{\mathrm{size}}{=}0.02$ assign secondary importance to efficiency.
A safety filter first removes any solution whose cross-validation
$F_1^w$ falls more than $\delta{=}0.003$ below the baseline $F_1^w$,
ensuring efficiency gains are never achieved at the cost of a
meaningful detection degradation.
The selected $(\tau^*,\theta^*)$ is used to refit the full pipeline on
the complete outer training set, and $M^*$ is evaluated exactly once
on the held-out test set.

In summary, the proposed LLM-MOO-AutoML framework
automates the full IDS pipeline in three coordinated stages.
A local Llama~3.1~8B model first analyzes a compact dataset
profile and generates a validated AutoDP and AutoFE policy
that narrows the NSGA-III search space to dataset-aware,
deployment-relevant bounds before any candidate is evaluated.
NSGA-III then jointly optimizes the LightGBM
feature-selection threshold and hyperparameters across three
objectives, weighted F1-score, P99 inference latency ratio,
and model size ratio, within the LLM-specified bounds,
producing a Pareto front of deployment-aware solutions in
50 trials with 3-fold cross-validation. Finally, the
deployment-oriented solution is selected from the Pareto
front using a utility score that prioritizes detection accuracy
while penalizing latency and model size, refitted on the full
training set, and evaluated once on the held-out test set.
The LLM guidance narrows the search space, while the MOO stage jointly optimizes weighted F1, P99 inference latency ratio, and model size ratio to identify IDS models for resource-constrained IoT and EVCS edge deployments. Algorithm~\ref{alg:llm_moo} summarizes the full procedure. 

\begin{algorithm}[!t]
  \caption{LLM-MOO-AutoML IDS}
  \label{alg:llm_moo}
  \begin{algorithmic}[1]
    \REQUIRE Dataset $\mathcal{D}$; LLM $\mathcal{M}$; policy space
             $\Pi$; folds $K{=}3$; trials $T{=}50$;
             weights $w_{F_1}{=}0.95$, $w_{P99}{=}0.03$,
             $w_{\mathrm{size}}{=}0.02$;
             F1-drop budget $\delta{=}0.003$
    \ENSURE $M^*$, Pareto front $\mathcal{F}$, $F^*$,
            $(\tau^*,\theta^*)$
    \STATE Split $\mathcal{D}\!\to\!\mathcal{D}_{\mathrm{tr}},
           \mathcal{D}_{\mathrm{te}}$
    \STATE $\mathcal{P}\leftarrow
           \{n,d,d_{\mathrm{num}},d_{\mathrm{cat}},\mathbf{c}_y\}$
           from $\mathcal{D}_{\mathrm{tr}}$
    \STATE $G\leftarrow\mathcal{M}(\mathcal{P},\,\text{temp}{=}0)$
    \STATE $\pi\leftarrow\mathrm{Validate}(G,\Pi)$
    \STATE Extract $\beta,\phi,\mathcal{T},\Theta$ from $\pi$
    \STATE Create $K$ folds; apply label encoding, z-score, AutoDP($\beta$),
           AutoFE($\phi$) per fold
    \STATE Train reference LightGBM; record $\{I_i\}$,
           $L^{\mathrm{base}}_{P99}$, $S^{\mathrm{base}}$,
           $F_1^{w,\mathrm{base}}$
    \STATE Initialise NSGA-III over $\mathcal{T}\times\Theta$
    \FOR{$t=1$ \TO $T{=}50$}
      \STATE Sample $(\tau_t,\theta_t)$;
             select $S_{\tau_t}$ via Eq.~(\ref{eq:autofs})
      \STATE Train LightGBM($\theta_t$); compute $F_1^w$,
             $R_{P99}$, $R_{\mathrm{size}}$ via $K$-fold CV
      \STATE Update NSGA-III population
    \ENDFOR
    \STATE Extract $\mathcal{F}$; compute HV
    \STATE Filter: keep $t\in\mathcal{F}$ with
           $F_1^{w,\mathrm{base}}-F_1^w\leq\delta$
    \STATE $(\tau^*,\theta^*)\leftarrow\arg\max\,u$ (Eq.~(\ref{eq:utility}))
    \STATE Refit pipeline on $\mathcal{D}_{\mathrm{tr}}$ with
           $(\tau^*,\theta^*)$
    \RETURN Evaluate $M^*$ on $\mathcal{D}_{\mathrm{te}}$
  \end{algorithmic}
\end{algorithm}

\section{Performance Evaluation}
\label{sec:eval}

\subsection{Experimental Setup}

All experiments use Python~3.14, Scikit-learn, LightGBM, Optuna (NSGA-III),
and Imbalanced-learn, with dataset-aware LLM policy bounds following the
Llama~3.1~8B framework described in Section~\ref{sec:framework}.
Fixed random seed:~42.

\begin{itemize}
  \item \textbf{CICIDS2017}~\citep{sharafaldin2018ids}: General network IDS
        benchmark; DoS, botnet, brute force, web attacks, port scan, and
        infiltration. Training subset: 48,569 samples, 77 features; test:
        566,149 samples.
  \item \textbf{IoTID20}~\citep{ullah2020iotid20}: IoT dataset with Mirai, DoS,
        scanning, and MITM. Representative subset: 62,578 samples (80/20
        split).
  \item \textbf{CICEVSE2024}~\citep{buedi2024evse}: EVCS dataset; 12 DoS,
        reconnaissance, and scanning attack types. Training: 47,338 samples,
        85 features; test: 548,940 samples.
\end{itemize}

For CICIDS2017 and CICEVSE2024, 2\% of training data is used during MOO
search; the full test set is used for final 
evaluation~\citep{yang2026moo_globecom}.
For IoTID20, a fixed 12,000-sample search subset with 3-fold CV is
used~\citep{yang2026llm_automl}. NSGA-III runs 50 trials with 3-fold CV.
Primary metric: weighted F1-score (due to class imbalance). Efficiency
metrics: P99 inference latency and model size.

\subsection{Baseline Methods}

Baseline LightGBM,
SOO-AutoML~\citep{yang2024autonomous_cyber}, MOO-AutoML~\citep{yang2026moo_globecom}, and
LLM-AutoML ~\citep{yang2026llm_automl} are reproduced under identical dataset partitions and evaluation protocol.

\subsection{Results and Analysis}

Tables~\ref{tab:cicids}-\ref{tab:evse} report detection performance and deployment efficiency metrics for all compared methods on CICIDS2017, IoTID20, and CICEVSE2024, respectively.

\begin{table}[H]
  \caption{Performance on CICIDS2017}
  \label{tab:cicids}
  \centering
  \scriptsize
  \renewcommand{\arraystretch}{1.1}
  \begin{tabular}{lcccc}
    \toprule
    \textbf{Method} & \textbf{F1 (\%)} & \textbf{P99 (ms)} & 
    \textbf{Size (MB)} & \textbf{Feat.} \\
    \midrule
    Baseline LightGBM                           
        & 98.914 & 0.0331 & 2.82  & 77 \\
    SOO-AutoML~\citep{yang2024autonomous_cyber} 
        & 99.172 & 0.0082 & 2.36  & 49 \\
    MOO-AutoML~\citep{yang2026moo_globecom}     
        & 99.200 & 0.0141 & 0.70  & 33 \\
    LLM-AutoML~\citep{yang2026llm_automl}       
        & 99.680 & 0.0198 & 2.02  & 30 \\
    \textbf{LLM-MOO-AutoML}                     
        & \textbf{99.788} & \textbf{0.0041} & 
          \textbf{1.2297} & \textbf{24} \\
    \bottomrule
  \end{tabular}
\end{table}

\begin{table}[H]
  \caption{Performance on IoTID20}
  \label{tab:iotid}
  \centering
  \scriptsize
  \renewcommand{\arraystretch}{1.1}
  \begin{tabular}{lcccc}
    \toprule
    \textbf{Method} & \textbf{F1 (\%)} & \textbf{P99 (ms)} &
    \textbf{Size (MB)} & \textbf{Feat.} \\
    \midrule
    Baseline LightGBM                        & 99.767 & 0.00369 & 0.3329 & 81 \\
    SOO-AutoML (Yang and Shami 2024)         & 99.075 & 0.00108 & 0.4818 & 59 \\
    MOO-AutoML (Yang 2026b)                  & 99.186 & 0.00124 & 0.4237 & 19 \\
    \textbf{LLM-MOO-AutoML}                  & \textbf{99.778} & \textbf{0.00038} & \textbf{0.1073} & \textbf{10} \\
    \bottomrule
  \end{tabular}
  \smallskip\\
  \raggedright\scriptsize
\end{table}

\begin{table}[H]
  \caption{Performance on CICEVSE2024}
  \label{tab:evse}
  \centering
  \scriptsize
  \renewcommand{\arraystretch}{1.1}
  \begin{tabular}{lcccc}
    \toprule
    \textbf{Method} & \textbf{F1 (\%)} & \textbf{P99 (ms)} & 
    \textbf{Size (MB)} & \textbf{Feat.} \\
    \midrule
    Baseline LightGBM                           
        & 99.651 & 0.0182 & 6.07  & 85 \\
    SOO-AutoML~\citep{yang2024autonomous_cyber} 
        & 99.973 & 0.1011 & 6.56  & 16 \\
    MOO-AutoML~\citep{yang2026moo_globecom}     
        & 99.981 & 0.03107 & 2.52  & 11 \\
    \textbf{LLM-MOO-AutoML}                     
        & \textbf{99.989} & \textbf{0.0057} & 
          \textbf{1.6308} & \textbf{10} \\
    \bottomrule
  \end{tabular}
\end{table}

\textbf{CICIDS2017.}
The proposed framework achieves 99.788\% weighted F1, outperforming
MOO-AutoML (99.200\%) and LLM-AutoML (99.680\%).
The LLM-guided AutoFE reduces the feature set from 77 to 24, smaller than
MOO-AutoML's 33, because the LLM narrows the importance-threshold range
before any NSGA-III trial is evaluated.
P99 latency (0.0041~ms) is competitive with MOO-AutoML (0.0141~ms) while
achieving a higher F1-score, and model size (1.2297~MB) remains compact
relative to most baselines.

\textbf{IoTID20.}
The proposed framework achieves 99.778\%
weighted F1, the highest among all compared methods,
while achieving the lowest P99 latency (0.00038~ms)
and smallest model size (0.1073~MB) among all
reproduced methods. The LLM AutoDP policy effectively
handles severe class imbalance without expanding the
training set unnecessarily, and the constrained MOO
search retains only 10 features, fewer than all
compared baselines.

\textbf{CICEVSE2024.}
The proposed framework achieves 99.989\% weighted F1, the highest among
all compared methods, while achieving the lowest P99 latency (0.0057~ms
vs.\ 0.03107~ms for MOO-AutoML) and smallest model size (1.6308~MB
vs.\ 2.52~MB for MOO-AutoML), retaining only 10 features.
The LLM policy narrows the feature-selection threshold range and
data-balancing strategy for this EVCS-specific dataset, reducing NSGA-III search time (672.0 s vs. 2276.5 s at 50 trials).

Across all three datasets, the LLM contribution is twofold: AutoDP guidance improves minority-class representation without unnecessary expansion; AutoFE guidance reduces the feature space before any MOO evaluation, so NSGA-III explores productive regions only. Compared with MOO-AutoML ~\citep{yang2026moo_globecom}, LLM guidance achieves higher weighted F1-scores across all three datasets (99.788\% vs. 99.200\% on CICIDS2017; 99.778\% vs. 99.186\% on IoTID20; 99.989\% vs. 99.981\% on CICEVSE2024) and retains fewer features (24 vs. 33 on CICIDS2017; 10 vs. 19 on IoTID20; 10 vs. 11 on CICEVSE2024), confirming that the LLM-narrowed search space focuses NSGA-III on more discriminative feature subsets. On IoTID20, the framework also achieves the lowest P99 latency (0.00038 ms vs. 0.00124 ms for MOO-AutoML) and smallest model size (0.1073 MB vs. 0.4237 MB), reflecting the joint optimization of weighted F1, P99 latency, and model size. These results confirm that LLM-guided search-space reduction and NSGA-III multi-objective optimization are complementary strategies that jointly improve detection performance and deployment efficiency across general networks, IoT, and EV charging environments.

\subsection{Search Efficiency Analysis}

Table~\ref{tab:searchtime} compares NSGA-III search time with and without LLM-guided bounds at five trial budgets across all three datasets. At the largest budget (50 trials), the LLM-guided configuration reduced search time by 34.0-70.5\% across the three datasets: on CICIDS2017 by 37.6\% (3535.0 s to 2206.4 s), on IoTID20 by 34.0\% (118.7 s to 78.3 s), and on CICEVSE2024 by 70.5\% (2276.5 s to 672.0 s). At smaller budgets, the guided search was faster in most comparisons. The one exception is CICIDS2017 at 10 trials, where it was 13.4\% slower (490.9 s vs. 432.9 s). On CICIDS2017 and CICEVSE2024, the reduction increases with the trial budget.

\begin{table}[H]
  \caption{NSGA-III Search Time (s) With vs.\ Without LLM Bounds}
  \label{tab:searchtime}
  \centering
  \scriptsize
  \renewcommand{\arraystretch}{1.1}
  \begin{tabular}{llccccc}
    \toprule
    \textbf{Dataset} & \textbf{Run} & \textbf{10} & \textbf{20} & 
    \textbf{30} & \textbf{40} & \textbf{50} \\
    \midrule
    \multirow{2}{*}{CICIDS2017}
      & No LLM   & 432.9  & 1068.9 & 1602.1 & 2261.6 & 3535.0 \\
      & With LLM & 490.9  &  924.6 & 1329.2 & 1778.5 & 2206.4 \\
    \midrule
    \multirow{2}{*}{IoTID20}
      & No LLM   &  16.5  &   43.7 &   63.2 &   87.4 &  118.7 \\
      & With LLM &  12.2  &   30.8 &   47.2 &   62.5 &   78.3 \\
    \midrule
    \multirow{2}{*}{CICEVSE2024}
      & No LLM   & 361.6  &  870.5 & 1266.1 & 1784.4 & 2276.5 \\
      & With LLM & 262.6  &  364.0 &  461.7 &  574.2 &  672.0 \\
    \bottomrule
  \end{tabular}
\end{table}

\section{Conclusion}
\label{sec:conclusion}

This paper proposed LLM-MOO-AutoML, a framework for efficient intrusion detection in general networks, IoT systems, and EVCSs. A local Llama 3.1 8B model generates bounded, validated AutoDP and AutoFE policies from a compact dataset profile, narrowing the NSGA-III search space to dataset-aware, deployment-relevant regions. NSGA-III then jointly optimizes the LightGBM feature-selection threshold and hyperparameters under three objectives: weighted F1-score, P99 inference latency ratio, and model size ratio. Experiments on CICIDS2017, IoTID20, and CICEVSE2024 show that the proposed framework achieves the highest weighted F1-scores on CICIDS2017 (99.788\%), IoTID20 (99.778\%), and CICEVSE2024 (99.989\%), with feature sets reduced to as few as 10 features, P99 latency as low as 0.00038 ms, and model size as small as 0.1073 MB. Compared with unconstrained multi-objective search, the LLM-guided configuration reduced NSGA-III search time by 34.0-70.5\% across the three datasets at the 50-trial budget, and in most comparisons at smaller budgets. These results indicate that dataset-aware search-space reduction shortens search time while maintaining deployment quality. Future work will investigate continual IDS learning under concept drift, evaluate additional local LLM sizes, and extend the framework to incorporate energy consumption as a fourth MOO objective for constrained edge deployments.

\section*{Acknowledgments}
This work was supported in part by the Department of National Defence (DND) and the Natural Sciences and Engineering Research Council of Canada (NSERC) through the DND/NSERC Discovery Grant Supplement under Grant DGDND-2025-05840, and in part by the Faculty of Business and Information Technology at Ontario Tech University through the Early Career Research Chair in Cybersecurity for Business program.

\bibliography{aaai2027}

\end{document}